\documentclass[showpacs,10pt,twocolumn,prb]{revtex4-1}
\usepackage{amsmath}
\usepackage{amssymb}
\usepackage{graphics}
\usepackage{epsfig}
\usepackage{CJK}

\begin{document}


\title{Pauli limited upper critical field in Fe$_{1+y}$Te$_{1-x}$Se$_{x}$}
\author{Hechang Lei,$^{1}$ Rongwei Hu,$^{1,\ast }$ E. S. Choi,$^{2}$ J. B.
Warren,$^{3}$ and C. Petrovic$^{1}$}

\affiliation{$^{1}$Condensed Matter Physics and Materials Science Department, Brookhaven
National Laboratory, Upton, NY 11973, USA}
\affiliation{$^{2}$NHMFL/Physics, Florida State University, Tallahassee, Florida 32310,
USA}
\affiliation{$^{3}$Instrumentation Division, Brookhaven National Laboratory, Upton, New
York 11973, USA}
\date{\today}

\begin{abstract}
In this work we investigated the temperature dependence of the upper
critical field $\mu _{0}H_{c2}(T)$ of Fe$_{1.02(3)}$Te$_{0.61(4)}$Se$%
_{0.39(4)}$ and Fe$_{1.05(3)}$Te$_{0.89(2)}$Se$_{0.11(2)}$ single crystals
by measuring the magnetotransport properties in stable dc magnetic fields up
to 35 T. Both crystals show that $\mu _{0}H_{c2}(T)$ in
the ab-plane and along the c-axis exhibit saturation at low temperatures. The anisotropy of $\mu
_{0}H_{c2}(T)$ decreases with decreasing temperature, becoming nearly
isotropic when the temperature T$\rightarrow$0. Furthermore, $\mu
_{0}H_{c2}(0)$ deviates from the conventional Werthamer-Helfand-Hohenberg
(WHH) theoretical prediction values for both field directions. Our analysis
indicates that the spin-paramagnetic pair-breaking effect is responsible for
the temperature-dependent behavior of $\mu _{0}H_{c2}(T)$ in both field
directions.
\end{abstract}

\maketitle

\section{Introduction}

The discovery of superconductivity in REOFePn (RE=rare earth; Pn=P or As,
1111-system) \cite{Kamihara}$^{-}$\cite{Wen HH} with high transition
temperature $T_{c}$ has generated a great deal of interests. Shortly after,
several other groups of iron-based superconductors have been discovered,
such as AFe$_{2}$As$_{2}$ (A=alkaline or alkaline-earth metals, 122-system),\cite{Rotter}$^{,}$\cite{Chen GF2} LiFeAs (111-system),\cite{Wang XC} (Sr$_{4}$M$_{2}$O$_{6}$)(Fe$_{2}$Pn$_{2}$) (M=Sc, Ti or V, 42622-system),\cite{Ogino}$^{,}$\cite{Zhu XY} and $\alpha $-PbO type FeSe (11-system).\cite{Hsu FC} In particular, the discovery of superconductivity in FeSe, FeTe$_{1-x}$Se$_{x}$,\cite{Yeh KW} and FeTe$_{1-x}$S$_{x}$\cite{Mizuguchi} opened new
directions. Simple binary Fe based superconductors can help to understand
the mechanism of superconductivity because they share the most prominent
characteristics with other iron-based superconductors, i.e., a square-planar
lattice of Fe with tetrahedral coordination and similar Fermi surface
topology.\cite{Subedi} Furthermore, 11-type superconductors exhibit some
distinctive features: absence of charge reservoir, significant pressure
effect,\cite{Mizuguchi2} and excess Fe with local moment.\cite{Zhang LJ}

In order to understand the mechanism of superconductivity of iron-based
superconductors, it is important to study the upper critical field $\mu
_{0}H_{c2}$. This is one of the most important superconducting parameters
since it provides valuable information on fundamental superconducting
properties: coherence length, anisotropy, details of underlying electronic
structures and dimensionality of superconductivity as well as insights into
the pair-breaking mechanism.

There are two remarkable common characteristics in $\mu _{0}H_{c2}$-$T$
phase diagram in iron-based superconductors. For 1111-system, $\mu
_{0}H_{c2,c}(T)$ shows pronounced upturn curvature at low temperatures. In
contrast, $\mu _{0}H_{c2,ab}(T)$ exhibits a downturn curvature with
decreasing temperature.\cite{Jaroszynski} The former can be explained by two
band theory with high diffusivity ratio of electron band to hole band and
the latter is mainly ascribed to the spin-paramagnetic effect.\cite{Hunte}$%
^{-}$\cite{Lee HS} For 122-system ((Ba, K)Fe$_{2}$As$_{2}$ and Sr(Fe, Co)$%
_{2}$As$_{2}$), the upturn curvature of $\mu _{0}H_{c2,c}(T)$ present in
1111-system does not appear, but it still shows positive curvature of
temperature far below T$_{c}$ without saturation. \cite{Yuan HQ}$^{-}$\cite%
{Altarawneh} It can also be interpreted using two band theory with smaller
diffusivity ratio of two bands when compared to 1111-system.\cite{Baily} On
the other hand $\mu _{0}H_{c2,ab}(T)$ tends to saturate with decreasing
temperature and it also originates from spin-paramagnetic effect.\cite{Kano} However, for KFe$_{2}$As$_{2}$, both $\mu _{0}H_{c2,ab}(T)$ and $\mu
_{0}H_{c2,c}(T)$ show saturation trend at low temperature with different negative curvature. The former can be ascribed to the spin-paramagnetic effect and the latter is mainly determined by orbital limited field in one band scenario.\cite{Terashima}

Previous research on polycrystalline FeSe$_{0.25}$Te$_{0.75}$ using pulsed
magnetic fields up to 55 T, indicated that spin-paramagnetic effect
dominates $\mu _{0}H_{c2}(T)$.\cite{Kida} However, it is necessary to
elucidate whether this kind of effect dominates the $\mu _{0}H_{c2,ab}(T)$ or
(and) $\mu _{0}H_{c2,c}(T)$. In this work, we report the upper critical
field anisotropy of Fe$_{1.02(3)}$Te$_{0.61(4)}$Se$_{0.39(4)}$ and Fe$%
_{1.05(3)}$Te$_{0.89(2)}$Se$_{0.11(2)}$ single crystals in stable dc high
magnetic field up to 35T. We show that spin-paramagnetic effect is dominant
in both of $\mu _{0}H_{c2,ab}(T)$ and $\mu _{0}H_{c2,c}(T)$.

\section{Experiment}

Single crystals of Fe(Te,Se) were grown by self-flux method with nominal
composition FeTe$_{0.5}$Se$_{0.5}$ and FeTe$_{0.9}$Se$_{0.1}$.
Stoichiometric elemental Fe (purity 99.98\%, Alfa Aesar ), Te (purity
99.999\%, Alfa Aesar), and Se (purity 99.999\%, Alfa Aesar) were sealed in
quartz tubes under partial argon atmosphere. The sealed ampoule was heated
to a soaking temperature of 950 $^\circ$C, then slowly cooled to 300-400 $^\circ$C. Plate-like crystals up to 2$\times $5$\times $1 mm$^{3}$ can be grown.
The powder X-ray diffraction (XRD) spectra were taken with Cu K$_{\alpha }$ radiation ($%
\lambda $=$1.5418$ \r{A}) using a Rigaku Miniflex X-ray machine. X-ray
diffraction (XRD) results of the ground samples indicate the phases for both
of them are pure. The lattice parameters, a=b=3.798(2) \r{A}, c=6.063(2) \r{%
A}\ and a=b=3.818(2) \r{A}, c=6.243(2) \r{A}\ for nominal composition FeTe$%
_{0.5}$Se$_{0.5}$ and FeTe$_{0.9}$Se$_{0.1}$, respectively, are obtained by
fitting the XRD spectra using the Rietica software.\cite{Hunter} On the
other hand, X-ray diffraction (XRD) spectra of single crystals reveal that
the crystal surface is normal to the c-axis with the plate-shaped surface
parallel to the ab-plane. The elemental and microstructure analysis were
performed using energy-dispersive x-ray spectroscopy in an JEOL JSM-6500
scanning electron microscope. The average stoichiometry was determined by
examination of multiple points on the crystals. The measured compositions
are Fe$_{1.02(3)}$Te$_{0.61(4)}$Se$_{0.39(4)}$ and Fe$_{1.05(3)}$Te$%
_{0.89(2)}$Se$_{0.11(2)}$. They will be denoted as Se-39 and
Se-11 in the following for brevity. Electrical transport measurements were
performed using a four-probe configuration on rectangular shaped polished
single crystals with current flowing in ab-plane of tetragonal structure.
Thin Pt wires were attached to electrical contacts made of Epotek H20E
silver epoxy. Sample dimensions were measured with an optical microscope
Nikon SMZ-800 with 10 $\mu$m resolution. Electrical transport measurements were
carried out in dc fields up to 9T in a Quantum Design PPMS-9 from 1.8 to 200
K and up to 35 T in a resistive magnet in a He3 cryostat down to 0.3 K at the
National High Magnetic Field Laboratory (NHMFL) in Tallahassee, FL.

\section{Results}

Temperature dependent resistivity of $\rho _{ab}(T)$ of Se-39 and Se-11
below 15 K in low magnetic fields from 0 to 9 T for H$\Vert $ab and H$\Vert $c are shown in Fig. 1. With increasing magnetic fields, the resistivity
transition width becomes slightly broader and the onset of superconductivity gradually shifts to lower temperatures. The trend is more pronounced for H$%
\Vert $c than H$\Vert $ab. This is similar to previous reports for Fe(Te,S) and FeTe$_{0.7}$Se$_{0.3}$ single crystals.\cite{Hu RW}$^{,}$\cite{Chen
GF3} It is worth noting that the shape and width of $\rho _{ab}(T)$
broadening with H$\Vert $c is comparable to that of the 122-system, e.g. the
single crystal of (Ba,K)Fe$_{2}$As$_{2}$ and (Ba,Rb)Fe$_{2}$As$_{2}$.\cite%
{Wang ZS}$^{,}$\cite{Bukowski} It is rather different from 1111-system such
as single crystal of SmO$_{0.7}$F$_{0.25}$FeAs and SmO$_{0.85}$FeAs.\cite%
{Lee HS}$^{,}$\cite{Karpinski} Similar field broadening of resistivity of
the 1111-system with H$\Vert $c was also observed in cuprates.\cite{Kwok}$%
^{-}$\cite{Safar} and explained by the vortex-liquid state.\cite{Blatter}
Recent report on NdFeAsO$_{1-x}$F$_{x}$ single crystals confirmed the
existence of the vortex-liquid state in 1111-system.\cite{Pribulova} Hence,
the vortex-liquid state region is narrower even absent in Fe(Te,Se)
(11-system).

\begin{figure}[tbp]
\centerline{\includegraphics[scale=0.8]{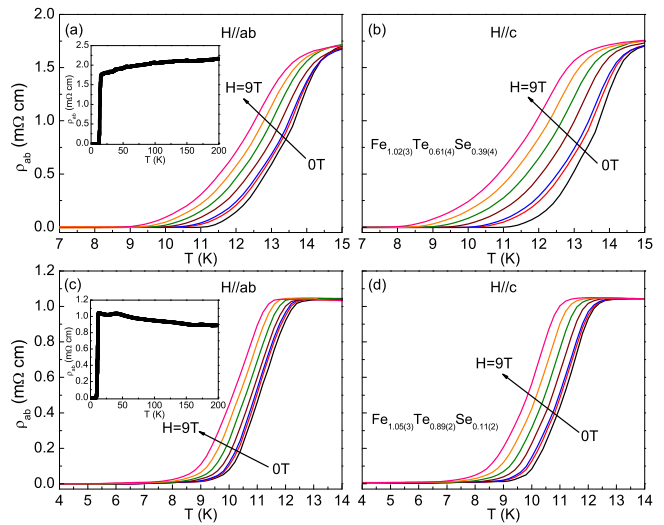}}
\vspace*{-0.3cm}
\caption{Temperature dependence of the
resistivity $\protect\rho _{ab}(T)$ of Se-39 for (a) H$\Vert
$ab and (b) H$\Vert $c of Se-39 and of Se-11 for (c) H$\Vert $ab and (d) H$\Vert $c at the
various magnetic fields from 0 to 9 T (0, 0.5, 1, 3, 5, 7, and 9T). Insets of (a) and (c) show the resistivity of Se-39 and Se-11 at the
temperature range of 1.8-200K, respectively.}
\end{figure}

Insets of Fig.\ 1(a) and (c) show the temperature dependence of the resistivity $\rho _{ab}(T)$ for Se-39 and Se-11 at zero field from 1.8K to 200 K. Both undergo a relatively sharp
superconducting transition at $T_{c,onset}$=14.4 K and 12 K for Se-39 and Se-11,
respectively. It should be noted that, as seen from the insets, Se-39
exhibits a metallic resistivity behavior in normal state, whereas Se-11 is
not metallic. This difference can be ascribed to different Se content.\cite%
{Yeh KW}$^{,}$\cite{Sales} The non-metallic behavior of Se-11 has also been
observed in low S doped FeTe single crystals.\cite{Hu RW} In addition, more
excess Fe in Fe(2) site for Se-11 than Se-39 could lead to weak charge
carrier localization.\cite{Chen GF3}$^{,}$\cite{Liu TJ} On the other hand, there is an anomalous peak in $\rho _{ab}(T)$ for Se-11 at T$\simeq $42K.
It corresponds to simultaneous structural and (or) antiferromagnetic transitions.
Comparing with undoped FeTe,\cite{Bao W} the transtion has been depressed from around
65K to 42K.

\begin{figure}[tbp]
\centerline{\includegraphics[scale=0.8]{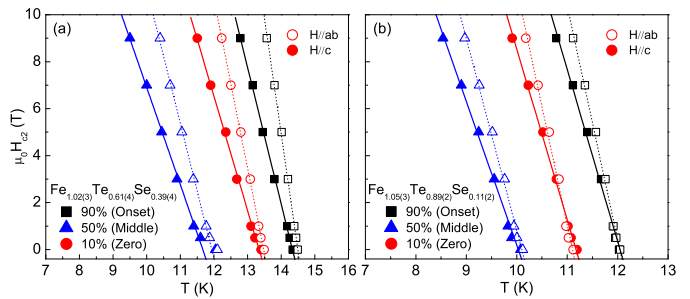}}
\vspace*{-0.3cm}
\caption{Temperature dependence of the
resistive upper critical field $\protect\mu _{0}H_{c2}(T)$ of (a) Se-39 and (b) Se-11 corresponding three defined temperatures at low fields. The
dotted and solid lines are linear fitting to the data for H$\Vert $c and H$%
\Vert $ab, respectively.}
\end{figure}

Fig. 2 shows the upper critical field $\mu _{0}H_{c2}(T)$ of Se-39 and Se-11
corresponding to the temperatures where the resistivity drops to 90\% of the
normal state resistivity $\rho _{n,ab}(T,H)(T_{c,onset})$, 50\% of $\rho
_{n,ab}(T,H)(T_{c,middle})$ and 10\% of $\rho _{n}(T,H)(T_{c,zero})$ in low
fields. The normal-state resistivity $\rho _{n,ab}(H,T)$ was determined by
linearly extrapolating the normal-state behavior above the onset of
superconductivity transition in $\rho _{ab}(T)$ curves (same as for $\rho _{ab}(H)$
curves). Because the curves of $\mu _{0}H_{c2}(T)$ for all defined
temperatures are almost linear except for $\mu _{0}H_{c2}(T_{c,zero})$ of
Se-39 with slightly upturn curvature near 0 T, we use the linear fitting
results at low field near $T_{c}$ as the slopes of $\mu _{0}H_{c2}(T_{c})$. This is shown by solid and dotted lines in Fig. 2 and the values are listed in Table
1. According to the conventional one-band Werthamer-Helfand-Hohenberg (WHH)
theory, which describes the orbital limited upper critical field of dirty type-II
superconductors,\cite{Werthamer} the $\mu _{0}H_{c2}^{\ast }(0)$ can be
described by
\begin{equation}
\mu _{0}H_{c2}^{\ast }(0)\text{=-0.693}(\frac{d\mu _{0}H_{c2}}{dT})_{Tc}Tc
\end{equation}%
and the values corresponding to three defined temperatures are also listed
in Table 1.

\begin{table*}[tbp] \centering

\caption{($d\mu _{0}H_{c2}/dT)_{Tc}$ and derived $\mu _{0}H_{c2}^{*}(0)$
data at three defined temperatures using WHH formula for
Se-39 and Se-11. $\mu _{0}H_{c2,ab}^{*}(0)$ and $\mu _{0}H_{c2,c}^{*}(0)$
are the ab-plane  and c-axis orbital limited upper critical fields at T=0K.}%
\begin{tabular}{ccccccc}
\hline\hline
&  & $T_{c}$ & $(d\mu _{0}H_{c2}/dT)_{Tc}$, H$\parallel $ab & $(d\mu
_{0}H_{c2}/dT)_{Tc}$, H$\parallel $c & $\mu _{0}H_{c2,ab}^{\ast }(0)$ & $\mu
_{0}H_{c2,c}^{\ast }(0)$ \\
&  & (K) & (T/K) & (T/K) & (T) & (T) \\ \hline
Fe$_{1.02(3)}$Te$_{0.61(4)}$Se$_{0.39(4)}$ & Onset & 14.4 & -9.9 & -5.8 &
98.8 & 57.9 \\
& Middle & 13.4 & -7.2 & -4.9 & 66.8 & 45.5 \\
& Zero & 12.1 & -5.7 & -4.1 & 47.8 & 34.4 \\ \hline
Fe$_{1.05(3)}$Te$_{0.89(2)}$Se$_{0.11(2)}$ & Onset & 12.0 & -10.0 & -7.1 &
83.1 & 59.0 \\
& Middle & 11.2 & -10.0 & -7.3 & 77.6 & 56.7 \\
& Zero & 10.1 & -8.2 & -6.1 & 57.4 & 42.7 \\ \hline\hline
\end{tabular}
\label{TableKey}
\end{table*}

The magnetic-field dependence of resistivity $\rho _{ab}(H)$ of Se-39 and
Se-11 are presented in
Fig. 3. It can be clearly seen that superconductivity is
suppressed by increasing magnetic field at the same temperature and the
transition of $\rho _{ab}(H)$ curves are shifted to lower magnetic fields at
higher measuring temperature. Comparing with Se-11, the superconductivity of
Se-39 still appears under field up to 35 T when temperature is below 1.47 K,
indicating Se-39 has a higher $\mu _{0}H_{c2}(0)$ than Se-11 in both
directions.

\begin{figure}[tbp]
\centerline{\includegraphics[scale=0.8]{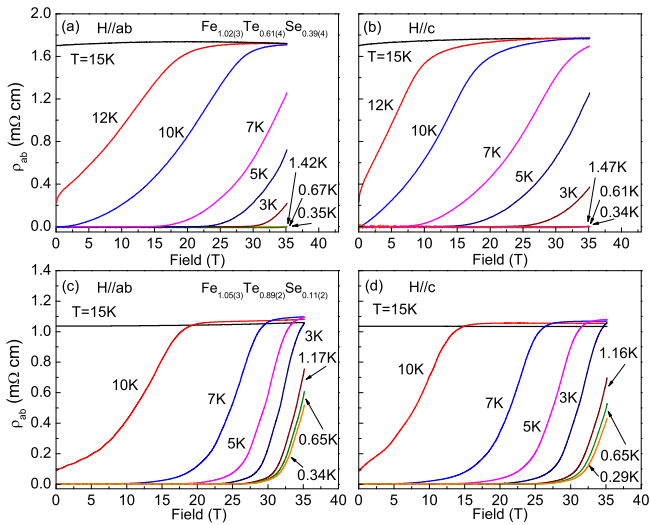}}
\vspace*{-0.3cm}
\caption{Field dependence of the
resistivity $\protect\rho _{ab}(H)$ of Se-39 for (a) H$\Vert $ab and
(b) H$\Vert $c, and of Se-11 for (c) H$\Vert $ab and (d) H$\Vert $c measured at
various temperatures in dc magnetic fields up to 35T.}
\end{figure}

Fig. 4 shows the temperature dependence of resistivity at high magnetic
fields. For Se-11, the superconductivity above 0.3 K is suppressed at $\mu
_{0}H$=35 T, irrespective of the direction of field. However it still
survives below 1.5 K for Se-39. This is consistent with the results of $\rho
_{ab}(H)$ measurement. The superconducting transition widths of both samples
are only slightly broader even at 35 T. It indicates that the vortex-liquid state in Fe(Te,Se) is much narrow or even absent in both low field high temperature region and high field low temperature region. On the other hand, the $\rho _{ab}(T)$ curves for H$\parallel $c and H$\parallel $ab
approach each other gradually with increasing filed. This trend is more
pronounced for Se-11 sample. The anisotropy of upper critical field is
decreasing with increasing field.

\begin{figure}[tbp]
\centerline{\includegraphics[scale=0.8]{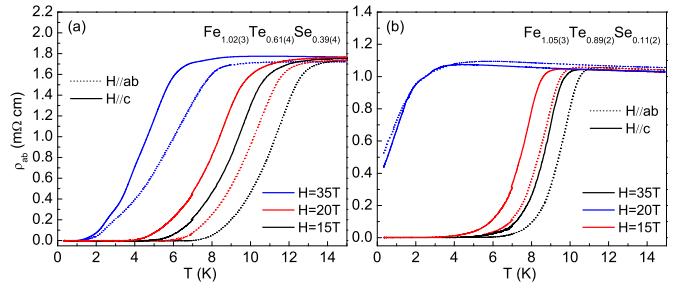}}
\vspace*{-0.3cm}
\caption{Temperature dependence of the resistivity $\protect\rho _{ab}$(T) at high magnetic fields from 15 to 35 T (15, 20,
and 35T) for (a) Se-39 and (b) Se-11.}
\end{figure}

By combining the magnetotransport results in low and high magnetic fields we
show phase diagrams in Fig. 5. Both samples show linear increase in $\mu
_{0}H_{c2}(T)$ with decreasing temperature near $T_{c}$. For Se-11 there is
a saturation trend at temperatures far below $T_{c}$ irrespective of field
direction. It can also be seen clearly that the $\mu _{0}H_{c2}(T)$ of Se-39
is higher than that of Se-11 for both field directions. Data above 35 T were
extracted by linear extrapolation of $\rho _{ab}(H)$ at $\mu _{0}H$ $<$ 35 T to $\rho _{ab}(H)$ =0.9$\rho _{n,ab}(T_{c},H)$. The
upper critical fields from high magnetic field measurement are much smaller
than those predicted using the conventional WHH model (Table 1), especially
for H$\Vert $ab.

\begin{figure}[tbp]
\centerline{\includegraphics[scale=0.8]{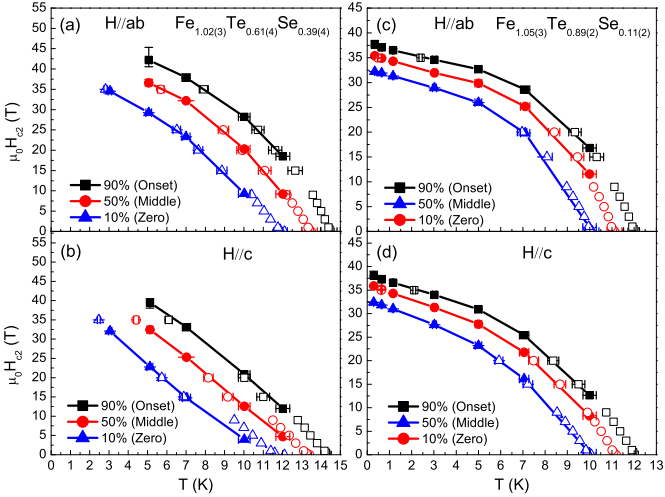}}
\vspace*{-0.3cm}
\caption{Temperature dependence of the
resistive upper critical field $\protect\mu _{0}H_{c2}(T)$ of Se-39 for (a) H$\Vert $%
ab and (b) H$\Vert $c and of Se-11 for (c) H$\Vert $ab and (d) H$\Vert $c obtained
from $\protect\rho _{ab}$(T) (open symbols) and $\protect\rho %
_{ab}$(H) (closed symbols) curves. Points above 35 T were extracted by
linear extrapolation of $\protect\rho _{ab}(H)$ at $\protect\mu _{0}H$ $<$ 35 T to $\protect\rho_{ab}(H)$=0.9$\protect\rho_{n,ab}(T_{c},H)$.}
\end{figure}

\section{Discussion}

In what follows we analyze the possible reasons for the deviation of $\mu
_{0}H_{c2}(0)$ from the conventional WHH model. Only the $\mu
_{0}H_{c2,onset}(T)$ were chosen for further analysis.\cite{Hunte}$^{,}
$\cite{Fuchs} In the conventional BCS model, orbital effect arising from the
Lorentz force acting on paired electrons with opposite momenta is the main
cause of pair breaking. The superconductivity is destroyed when the kinetic
energy exceeds the condensation energy of the Cooper pairs. On the other
hand, superconductivity can also be eliminated via breaking the singlet pair
into unbound triplet. In other words, the Pauli spin susceptibility energy
exceeding the condensation energy leads to the partial alignment of the
spins. This is spin Zeeman effect, also called spin-paramagnetic effect. The
effects of Pauli spin paramagnetism and spin-orbit interaction were included
in the WHH\ theory through the Maki parameters $\alpha $ and $\lambda _{so}$.%
\cite{Maki} For an isotropic type-II superconductor in the dirty limit, $\mu
_{0}H_{c2}(T)$ can be calculated using the following equation in terms of
digamma functions:\cite{Werthamer}

\begin{multline}
\ln \frac{1}{t}\text{=}(\frac{1}{2}+\frac{i\lambda _{so}}{4\gamma })\psi (%
\frac{1}{2}+\frac{\overset{-}{h}+\lambda _{so}/2+i\gamma }{2t}) \\
+(\frac{1}{2}-\frac{i\lambda _{so}}{4\gamma })\psi (\frac{1}{2}+\frac{%
\overset{-}{h}+\lambda _{so}/2-i\gamma }{2t})-\psi (\frac{1}{2})
\end{multline}

where $t$=$T/T_{c}$, $\gamma \equiv \lbrack (\alpha \overset{-}{h}%
)^{2}-(\lambda _{so}/2)^{2}]^{1/2}$ and

\begin{equation}
h^{\ast }\equiv \frac{\overset{-}{h}}{(-d\overset{-}{h}/dt)_{t=1}}\text{=}%
\frac{\pi ^{2}\overset{-}{h}}{4}\text{=}\frac{H_{c2}}{(-dH_{c2}/dt)_{t=1}}
\end{equation}

Here, we assume that $\lambda _{so}$=$0$ because the spin-orbit scattering
is expected to be rather weak\cite{Fuchs} and the equation can be simplified
as:%
\begin{equation}
\ln \frac{1}{t}\text{=}\frac{1}{2}\psi (\frac{1}{2}+\frac{(1+\alpha )\overset%
{-}{h}}{2t})+\frac{1}{2}\psi (\frac{1}{2}+\frac{(1-\alpha )\overset{-}{h}}{2t%
})-\psi (\frac{1}{2})
\end{equation}

When $\alpha $=$0$, in the absence of the spin-paramagnetic effect and the
spin-orbit interaction, orbital limited upper critical field $H_{c2}^{\ast }$
is described by:%
\begin{equation}
\ln \frac{1}{t}\text{=}\psi (\frac{1}{2}+\frac{\overset{-}{h}}{2t})-\psi (%
\frac{1}{2})
\end{equation}

and $\mu _{0}H_{c2}^{\ast }(0)$=-0.693$(d\mu _{0}H_{c2}/dT)_{T_{c}}T_{c}$, i.e., eq.
(1).

As shown in the Fig. 6, the data points of $\mu _{0}H_{c2}(T)$ for H$\Vert $%
ab and H$\Vert $c in both samples cannot be explained well using the WHH
model with $\alpha $=0 and $\lambda _{so}$=0 (Fig. 6(a,b) solid lines). We
obtain excellent fits for the $\mu _{0}H_{c2,ab}(T)$ and $\mu _{0}H_{c2,c}(T)
$ in Fig 6(a,b) using eq (4) with spin-paramagnetic effect. These results
indicate that the spin-paramagnetic effect is the dominant pair-breaking
mechanism in Se-39 and Se-11 for both H$\Vert $ab and H$\Vert $c.

\begin{figure}[tbp]
\centerline{\includegraphics[scale=0.8]{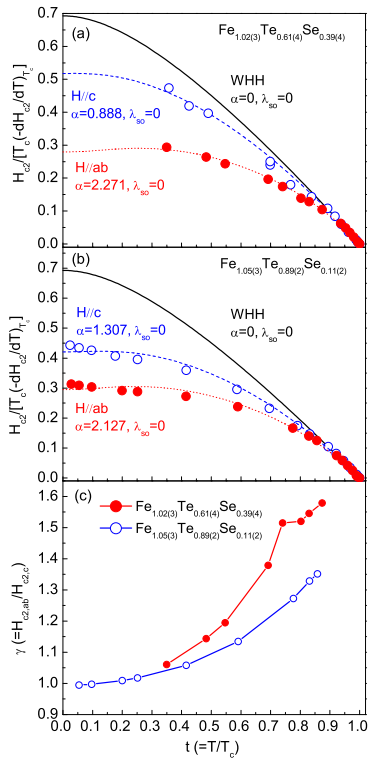}}
\vspace*{-0.3cm}
\caption{Normalized upper critical field
$h^{\ast }$ vs. reduced temperature $t$=$T/T_{c}$ for (a) Se-39 and (b) Se-11 for H$\Vert $ab (closed circle) and H$\Vert $c (open circle). Solid lines: WHH model with $\protect\alpha $=0, $\protect\lambda $=0; Dotted and dash
lines: fitted $h^{\ast }(t)$ including spin-paramagnetic effect for
H$\Vert $ab and H$\Vert $c, respectively. (c) The anisotropy in the upper
critical field, $\protect\gamma $=$H_{c2,ab}(T)/H_{c2,c}(T)$, as a function
of reduced temperature $t$=$T/T_{c}$.}
\end{figure}

The paramagnetically limited field $\mu _{0}H_{c2}^{p}(0)$ is given by $\mu
_{0}H_{c2}^{p}(0)$=$\mu _{0}H_{c2}^{\ast }(0)$/$\sqrt{1+\alpha ^{2}}$ and $%
\alpha $=$\sqrt{2}H_{c2}^{\ast }(0)$/$H_{p}(0)$, where $\mu _{0}H_{p}(0)$ is
zero-temperature Pauli limited field.\cite{Maki} The calculated $\mu
_{0}H_{c2}^{p}(0)$ and $\mu _{0}H_{p}(0)$ using $\alpha $ obtained from $%
H_{c2}(T)$ data fitting are listed in Table 2. From the $\mu _{0}H_{c2}(0)$
zero-temperature coherence length $\xi (0)$ can be estimated with
Ginzburg-Landau formula $\mu _{0}H_{c2}(0)$=$\Phi _{0}$/2$\pi \xi^{2}(0)$,
where $\Phi _{0}$=2.07$\times $10$^{-15}Wb$ (Table 2). The $\mu _{0}H_{c2}(0)
$ (determined by $\mu _{0}H_{c2}^{p}(0)$) of Se-39 in both field directions
are close to previously reported.\cite{Fang MH} Our results suggest that
Fe(Te,Se) exhibits the spin-singlet pairing in the superconducting state.
One the other hand, we also analyze our data using the two-band theory\cite%
{Jaroszynski}$^{,}$\cite{Gurevich}, and the fits are unsatisfactory the
two-band model (not shown here).

\begin{table*}[tbp] \centering%

\caption{Superconducting parameters of Se-39 and Se-11 obtained from the analysis of $\mu_{0}H_{c2,onset}(T)$.
 $\mu_{0}H_{c2}^{*}(0)$, $\mu_{0}H_{c2}^{p}(0)$ and $\mu_{0}H_{p}(0)$ are the zero-temperature orbital, paramagnetically, and Pauli limited upper critical fields, respectively. $\alpha$ is the fitted Maki parameter
    ($\lambda_{so}$=0). $\xi_{ab}(0)$ and $\xi_{c}(0)$ are the ab-plane and c-axis zero-temperature coherence length calculated using $\mu_{0}H_{c2}^{p}(0)$, respectively.}%
\begin{tabular}{ccccccccccc}
\hline\hline
& $\mu _{0}H_{c2,ab}^{\ast }(0)$ & $\mu _{0}H_{c2,c}^{\ast }(0)$ & $\mu
_{0}H_{c2,ab}^{p}(0)$ & $\mu _{0}H_{c2,c}^{p}(0)$ & $\mu _{0}H_{p,ab}(0)$ & $%
\mu _{0}H_{p,c}(0)$ & $\alpha _{H\Vert ab}$ & $\alpha _{H\Vert c}$ & $\xi
_{ab}(0)$ & $\xi _{c}(0)$ \\
& (T) & (T) & (T) & (T) & (T) & (T) &  &  & (nm) & (nm) \\ \hline
Fe$_{1.02(3)}$Te$_{0.61(4)}$Se$_{0.39(4)}$ & 98.8 & 57.9 & 39.8 & 43.3 & 61.5
& 92.2 & 2.271 & 0.888 & 2.76 & 3.00 \\
Fe$_{1.05(3)}$Te$_{0.89(2)}$Se$_{0.11(2)}$ & 83.1 & 59.0 & 35.4 & 35.9 & 55.3
& 63.8 & 2.127 & 1.307 & 3.03 & 3.07 \\ \hline\hline
\end{tabular}%
\label{TableKey copy(1)}
\end{table*}

It is instructive to discuss the origin of enhancement of spin-paramagnetic
effect, i.e., reduced values of $\mu _{0}H_{p}(0)$. The Maki parameter $%
\alpha $ is enhanced for disordered systems.\cite{Fuchs}$^{,}$\cite{Fuchs2}
For Se-39, more Se doping introduces more disorder than in Se-11. This
effect could contribute to larger $\alpha _{H\Vert ab}$ of Se-39 when
compared to Se-11. However, it cannot explain the inverse trend of $\alpha
_{H\Vert c}$. Therefore another effect must compete with disorder. This may
be the effect of excess Fe in Fe(2) position.

Excess Fe in Fe(2) position is the unique feature of 11-system, different
from other Fe pnictide superconductors. The Fe(2) has larger local magnetic
moment than Fe(1) in Fe-(Te,Se) layers. The Fe(2) moment is present even if
the SDW antiferromagnetic ordering of the Fe plane is suppressed by doping
or pressure, contributing to $N(E_{F})$.\cite{Zhang LJ} According to the
expression of $\mu _{0}H_{p}(0)$ with strong coupling correction considering
e-boson and e-e interaction:\cite{Fuchs}$^{,}$\cite{Orlando}$^{,}$\cite%
{Schossmann}

\begin{equation}
\mu _{0}H_{p}(0)\text{=}1.86(1+\lambda )^{\varepsilon }\eta _{\Delta }\eta
_{ib}(1-I)
\end{equation}

where $\eta _{\Delta }$ describes the strong coupling intraband correction
for the gap, $I$ is the Stoner factor $I$=$N(E_{F})J$, $N(E_{F})$ is the
electronic density of states (DOS) per spin at the Fermi level $E_{F}$, $J$
is an effective exchange integral, $\eta _{ib}$ is introduced to describe
phenomenologically the effect of the gap anisotropy, $\lambda $ is electron
-- boson coupling constant and $\varepsilon $=0.5 or 1. It can be seen that $%
\mu _{0}H_{p}(0)$ can decrease if the Stoner factor increases via
enhancement of $J$ or $N(E_{F})$. Excess Fe in Fe(2) site with local
magnetic moment could interact with itinerant electron in Fe layer,
resulting in exchange enhanced Pauli paramagnetism or
Ruderman-Kittel-Kasuya-Yosida (RKKY) interaction, thus enhancing $J$. Hence,
higher content of excess Fe in Se-11, could lead to larger $\alpha _{H\Vert
c}$ than in Se-39. Another possibility may be that the $N(E_{F})$ is
decreased with increasing the content of Se.\cite{Subedi} This trend will also enhance the Pauli limited field, i.e. suppress the spin-paramagnetic effect, according to above formula. This could be why the $\mu _{0}H_{p}(0)$ of Se-39 is higher than that of
Se-11 if we assume other parameters in eq. (6) are not changed.

Finally, we discuss the anisotropy of $\mu _{0}H_{c2}(T)$. The temperature
dependence of anisotropy of $\mu _{0}H_{c2}(T)$, $\gamma ($=$%
H_{c2,ab}(T)/H_{c2,c}(T))$, obtained from the $\mu _{0}H_{c2,onset}(T)$ data
is shown in Fig. 6(c) as a function of reduced temperature $t$=$T/T_{c}$.
The $\gamma $ of Se-11 is smaller than that of Se-39. The difference in $%
\gamma $ between the two samples decreases gradually. Both $\gamma $ values
decrease to about 1 with decreasing temperature, larger than in Fe(Te,S) and
similar to previously reported in Fe(Te,Se).\cite{Hu RW}$^{,}$\cite{Fang MH}
These results show that Fe(Te,Se) is a high-field isotropic superconductor.

\section{Conclusion}

In summary, the anisotropy in the upper critical field of Fe$_{1.02(3)}$Te$%
_{0.61(4)}$Se$_{0.39(4)}$ and Fe$_{1.05(3)}$Te$_{0.89(2)}$Se$_{0.11(2)}$
single crystals was studied in high and stable magnetic fields up to 35 T.\
It is found that the zero-temperature upper critical field is much smaller
than the predicted result of WHH theory without the spin-paramagnetic
effect. The anisotropy of the upper critical field decreases with decreasing
temperature, becoming nearly isotropic at low temperature. The
spin-paramagnetic effect is the dominant pair-breaking mechanism for both of
H$\Vert $ab and H$\Vert $c.

\section{Acknowledgements}

We thank T. P. Murphy for useful discussions and experiment support in
NHMFL. This work was carried out at the Brookhaven National Laboratory,
which is operated for the U.S. Department of Energy by Brookhaven Science
Associates DE-Ac02-98CH10886. This work was in part supported by the U.S.
Department of Energy, Office of Science, Office of Basic Energy Sciences as
part of the Energy Frontier Research Center (EFRC), Center for Emergent
Superconductivity (CES). A portion of this work was performed at the
National High Magnetic Field Laboratory, which is supported by NSF
Cooperative Agreement No. DMR-0084173, by the State of Florida, and by the
U.S. Department of Energy.

$^{\ast }$Present address: Ames Laboratory US DOE and Department of Physics
and Astronomy, Iowa State University, Ames, IA 50011, USA.

\end{document}